\documentclass[graybox]{svmult}

\usepackage{mathptmx}       
\usepackage{helvet}         
\usepackage{courier}        
\usepackage{type1cm}        
\usepackage{makeidx}         
\usepackage{graphicx}        
\usepackage{multicol}        
\usepackage[bottom]{footmisc}

\makeindex             

\begin{document}

\title*{Are there age spreads in star forming regions?}
\author{R. D. Jeffries}
\institute{R. D. Jeffries \at Astrophysics Group, Keele University,
  Staffordshire, ST5 5BG, UK, \email{rdj@astro.keele.ac.uk}
}
%
%
\maketitle

\abstract*{A luminosity spread at a given effective temperature is
ubiquitously seen in the Hertzsprung-Russell (HR) diagrams of young
star forming regions and often interpreted in terms of a prolonged
period ($\geq 10$\,Myr) of star formation. I review the evidence that the
observed luminosity spreads are genuine and not caused by astrophysical
sources of scatter. I then address whether the luminosity spreads
necessarily imply
large age spreads, by comparing HR diagram ages with ages from
independent clocks such as stellar rotation rate, the presence of
circumstellar material and lithium depletion. I argue that whilst there
probably is a true luminosity dispersion, there is little evidence to
support age spreads larger than a few Myr. This paradox could be
resolved by brief periods of rapid accretion during the class I pre main-sequence
phase.}

\abstract{A luminosity spread at a given effective temperature is
ubiquitously seen in the Hertzsprung-Russell (HR) diagrams of young
star forming regions and often interpreted in terms of a prolonged
period ($\geq 10$\,Myr) of star formation. I review the evidence that the
observed luminosity spreads are genuine and not caused by astrophysical
sources of scatter. I then address whether the luminosity spreads
necessarily imply
large age spreads, by comparing HR diagram ages with ages from
independent clocks such as stellar rotation rate, the presence of
circumstellar material and lithium depletion. I argue that whilst there
probably is a true luminosity dispersion, there is little evidence to
support age spreads larger than a few Myr. This paradox could be
resolved by brief periods of rapid accretion during the class I pre main-sequence
phase.}

\section{Introduction}
\label{intro}
When newly born stars emerge from their natal clouds as class II and
class III pre main-sequence (PMS) objects, they can be placed in a
Hertzsprung-Russell (HR) diagram. Low-mass ($<2\,M_{\odot}$)
stars take 10--100\,Myr to descend the Hayashi track
and settle onto the zero-age main-sequence, so the HR diagram can
be used, in combination with theoretical models, to estimate individual
ages for PMS stars or construct the age distribution of a group of
PMS stars.  The HR diagrams of young star forming regions (SFRs)
usually have an order of magnitude range of luminosity at a given
effective temperature ($T_{\rm eff}$, see Fig.1), and this luminosity dispersion is often
interpreted as star formation that has been ongoing for $\geq 10$\,Myr 
within a single SFR or young cluster (e.g. for young, nearby SFRs -- 
Palla \& Stahler 1999, 2000; for massive young clusters -- Beccari et
al. 2010; or even for resolved star clusters in other galaxies -- Da
Rio, Gouliermis \& Gennaro 2010a).

The presence and extent of any age spread is an important constraint on
models of star formation. A significant ($\geq 10$\,Myr) spread would
favour a ``slow'' mode, where global collapse is impeded by, for
example, a strong magnetic field (e.g. Tassis \& Mouschovias 2004). 
Age spreads that were $\leq 1$\,Myr however,
could be explained by the rapid dissipation of turbulence and star
formation on a dynamical timescale (e.g. Elmegreen 2000). The reality
or not of age spreads is also important from a practical point of
view. Ages from the HR diagram are used to understand the progression
of star formation (e.g. triggering scenarios, collect-and-collapse
models) and the age-dependent masses estimated from an HR diagram are
usually the only way of determining the initial mass function.
In this short review, I ask: 
\begin{enumerate}
\item Are the luminosity spreads (at a given $T_{\rm eff}$) in the HR diagram real? 

\item If so, do these necessarily imply a wide spread of ages within an individual SFR?
\end{enumerate}

\section{Luminosity spreads?}
\label{lumspread}

Hartmann (2001) identified many sources of astrophysical and
observational scatter that contribute to an {\em apparent} spread
in the luminosities of PMS stars at a given $T_{\rm eff}$. These
include the likelihood that many ``stars'' are unresolved multiples;
that individual stars may be subject to a range of
extinction and reddening; that PMS stars can be
highly variable; that the luminosity contributed by
accretion processes could vary from star-to-star; that in (nearby) SFRs
the stars are at a range of distances; and that placing stars on a
HR diagram requires temperature (or spectral type
or colour) and luminosity (brightness) measurements which have
observational uncertainties. Hartmann concluded that efforts to infer
star formation histories would be severely hampered by these effects
and that the luminosity and hence age spreads claimed by Palla \&
Stahler (2000), among others, must be extreme upper limits.
Hillenbrand, Bauermeister \& White (2008)
showed that it is difficult to verify or indeed quantify luminosity
spreads, and hence infer age spreads, unless (a) observational
uncertainties are small and (b) both the {\it size and distribution} of
other astrophysical sources of luminosity dispersion are well
understood.

One approach to tackle these difficulties is to quantify
spreads that could be contributed by individual sources of dispersion
and model the outcome.  Burningham et al. (2005) used photometric
measurements at more than one epoch to empirically assess the affects
of variability on two young SFRs ($\sigma$~ Ori and Cep OB3b) with
significant (compared to observational uncertainties) scatter in their
colour-magnitude diagrams (CMDs). This approach takes account of
correlated variability in colours and magnitudes and the non-Gaussian
distribution of variability-induced dispersion. A coeval population was
simulated using the observed levels of variability, the likely
effects of binarity and observational errors. This model was found to
significantly {\it underpredict} the observed dispersion. In other
words, variability (on timescales of years or less), binarity and
observational error could only account for a small fraction of the
luminosity dispersion. On the other hand, Slesnick, Hillenbrand \&
Carpenter (2008) examined the slightly older Upper Sco SFR and showed
that the large observed luminosity spreads could perhaps be entirely
explained by a coeval population affected by a combination of
observation errors, distance dispersion and binarity. However, the
additional dispersion (particularly due to distance uncertainties) was
so large in this case that additional scatter equivalent to a real age
dispersion of $\pm 3$\,Myr remained a possibility.

A more sophisticated statistical approach has been taken by Da Rio,
Gouliermis \& Gennaro (2010a) who, using a maximum likelihood method
akin to that proposed by Naylor \& Jeffries (2006), fitted a
2-dimensional synthetic surface density to the CMD of a SFR in the
Large Magellanic Cloud. The model includes contributions from
unresolved binarity, variability, differential extinction and
accretion. These authors conclude that the luminosity spread in the CMD
is too large to be accounted for by the ``nuisance'' sources of
dispersion and interpret the additional scatter as a spread in ages of
FWHM 2.8--4.4 Myr

An alternative for investigating the reality of the luminosity
dispersions is to examine proxies such as radius or gravity that would
be expected to show a corresponding dispersion, but whose measurement
is not so greatly affected by the additional astrophysical sources of
scatter. An example is the use of rotation periods and projected
equatorial velocities to estimate the projected radii, $R \sin i$, of
PMS stars in the Orion Nebula cluster (ONC, Jeffries 2007). These
measurements are largely unaffected by binarity, variability,
differential extinction, distance or accretion. Assuming that spin-axes
are randomly oriented, the distribution of $R \sin i$ can be modelled
to estimate mean radii and the extent of any true spread in radius at a
given $T_{\rm eff}$. The results confirm that a factor of 2--3 (FWHM)
spread in radius exists at a given $T_{\rm eff}$ and this concurs with
the order of magnitude luminosity spread seen in the HR diagram of the
same objects.

In summary, although there are few detailed investigations to draw on,
the evidence so far suggests that the luminosity spreads seen in
SFRs are mostly genuine. Only a fraction of the dispersion can be
explained by observational uncertainties, variability, binarity and
accretion.

\section{Age Spreads?}
\label{agespread}

If the luminosity dispersions are genuine, then it is natural to plot a
set of HR diagram isochrones, estimate an age for each star and hence
infer an age distribution. However it is possible that physical causes
other than age could contribute to a real dispersion of luminosity in
the HR diagram of young PMS stars. Accretion could perturb the
evolution of the central star, inducing a luminosity spread even in a
coeval population (Tout, Livio \& Bonnell 1999). To investigate the
fidelity of ages deduced from the HR diagram we can compare these ages
with those estimated using independent clocks. These include the
depletion of photospheric lithium, the evolution of stellar rotation
and the dispersal of circumstellar material.

\subsection{Lithium Depletion} %
\label{li}

Lithium is ephemeral in the photospheres of young, low-mass stars. Once
the central temperature of a star reaches the Li ignition temperature,
($\sim 2.5\times 10^{6}$\,K) convective mixing leads to almost
complete Li depletion unless the PMS star leaves the Hayashi track and
develops a radiative core (see Jeffries 2006). In principle the level
of Li in the atmosphere of a low-mass PMS star is a mass-dependent
clock. Palla et al. (2005) and Sacco et al. (2007) have searched for
Li-depleted stars that are bona-fide members of the Orion Nebula
cluster and the $\sigma$~Ori and $\lambda$~Ori associations. They do
find a few such objects (a few per cent of the total) and using models for Li depletion,
infer ages for them of $>10$\,Myr, compared to HR diagram ages
of 2--5\,Myr for the bulk of the PMS population. These observations
are consistent with the presence of a small fraction of older objects,
co-existing with the bulk of the younger PMS population, arguing in
favour of a large age spread. 

Whilst this interpretation is possible, there are some problems. First,
the bimodal distribution of Li abundances (i.e. most stars are
undepleted with a small fraction of extremely Li-depleted objects) does
not seem consistent with a smooth underlying distribution of ages and
indeed contamination by older, non-members of the cluster has been
suggested (Pflamm-Altenburg \& Kroupa 2007). Second, although in some
(but not all) cases, the Li-depletion age for these stars matches the
HR diagram age, they are {\it not} fully independent age
indicators. The central temperature of the star, which controls the
Li-burning, will depend on the stellar radius (and hence luminosity in
the HR diagram). If for some reason the star had a smaller radius than
expected at a given age and therefore appeared older in the HR
diagram, its central temperature would {\it also} be higher and it would have
a greater capacity to burn Li.

\subsection{Rotation rates}
\label{rotation}

Young, PMS stars typically rotate with periods of 1--10\,d. There is
strong evidence that PMS stars with circumstellar disks and active
accretion rotate more slowly on average than those without disks
(e.g. Rebull et al. 2006; Cieza \& Baliber 2007). A widely accepted
idea is that stars which are accreting from a disk are braked by the
star-disk interaction and held at a roughly constant spin period
(Rebull, Wolff \& Strom 2004). Once the disk disperses, or below some
threshold accretion rate, the brake is released and the star spins up
as it rapidly contracts along the Hayashi track. Thus, the rotation
rate of PMS stars should broadly reflect the age of the population --
an older population should have fewer strong accretors (see
section~\ref{accrete}), have had more time to spin-up, and hence should
contain a greater proportion of fast rotators than a younger
population.  As the lifetime of accretion is of order a few Myr, then
age spreads of 10~Myr should manifest themselves as big differences in
the rotation period distributions of the ``older'' and ``younger''
populations.

This rotation clock has been investigated by Littlefair et
al. (2011). They divided the PMS populations of several nearby SFRs
into ``old'' (low luminosity) and ``young'' (high luminosity) samples
and compared their rotation period distributions. The null hypothesis
that the samples were drawn from the same distribution could be
rejected at high significance levels,  but the surprising result is
that the faster rotating sample is actually the one containing the
``young'' objects. If the luminosity spreads were truly caused by an
age spread, the ``disk-locking'' model would predict the opposite result.
Littlefair et al. interpret this by assuming the populations in each
SFR are coeval, but the luminosity spreads are introduced through
differing accretion histories which also influence the stellar rotation
rate (see section~\ref{interpret}).

\subsection{Disk dispersal}
\label{accrete}

%
\begin{figure}[t]
\includegraphics[scale=.45]{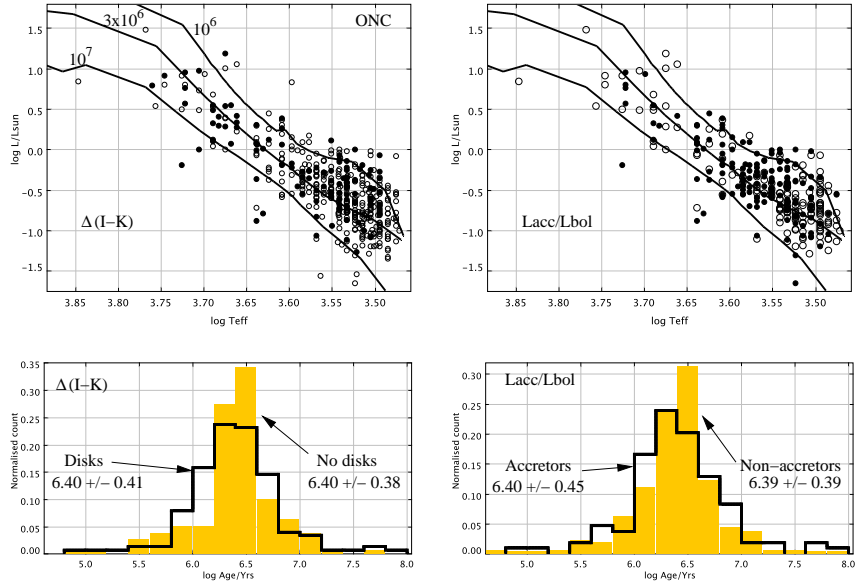}
%
%
\caption{The HR diagrams and inferred age distributions for samples of
  stars in the Orion Nebula cluster (ONC, data from Da Rio et al. 2010b).
  (Left) Upper plot shows isochrones (from Siess et al. 2000, labelled
  in Myr) and
  stars in the ONC separated by infrared excess. Open symbols are stars
  with $\Delta (I-K)>0.3$ (data from Hillenbrand et al. 1998). Lower
  diagram shows the age distributions which have identical means and
  similar dispersions. (Right) A similar plot, but the open symbols are
  stars with $L_{\rm accrete}/L_{\rm bol}>0.1$ (from Da Rio et
  al. 2010b). Again, the lower plot shows the age distributions of
  these samples are very similar.
}
\label{fig1}       
\end{figure}

It is well known that the lifetime of circumstellar material around
young PMS stars, traced by the fraction of objects exhibiting
infrared excesses or accretion diagnostics, is on average a few
Myr (e.g. Haisch, Lada \& Lada 2001; Calvet et al. 2005; Jeffries et
al. 2007; Hern\'andez et al. 2008).  The precise reasons for disk
dispersal are still unclear, but if the fraction of stars
accreting strongly from a circumstellar disk does decrease with age
then we would expect to see fewer active accretors among any older
population {\it within a single SFR}. 

Surprisingly little work has been done in this area. Hartmann et
al. (1998) found that mass accretion rates did decline with increasing HR
diagram age in Taurus and Chamaeleon. Bertout, Siess \& Cabrit (2007)
claimed that accreting classical T-Tauri stars in Taurus appeared significantly
younger in the HR diagram than their weak-lined, non-accreting
counterparts. On the other hand, Hillenbrand et al. (1998) find no
correlation between age and the fraction of PMS stars in the ONC with
near-infrared excesses. These studies are difficult because they are
afflicted by a number of biases and selection effects. 

In preparation for this review I examined a new catalogue of sources in
the ONC by Da Rio et al. (2010b), which they claim to be complete to
very low luminosities. They have estimated the luminosity and effective
temperature of stars using a careful star-by-star estimate of accretion
luminosity and extinction. Their catalogues give estimated masses and
ages based on the models of Siess, Dufour \& Forestini (2000). Figure~1
shows HR diagrams and deduced age distributions, where the samples have
been divided according to (a) whether the $I-K$ excess over a
photospheric colour is $>0.3$ (data from Hillenbrand et al. 1998) or
(b) whether the accretion luminosity is $>0.1\,L_{\rm bol}$. Neither of
these accretion/disk diagnostics shows a significant age dependence
within the ONC, the mean ages and age distributions of the subsamples
are indistinguishable. I am currently exploring any possible biases
(e.g. dependences of age and the likelihood of possessing a disk on
position within the cluster) that might explain these results.
 
Taking the results at face value suggests either: (i) Any true age
spreads are much less than the few Myr characteristic timescale for the
cessation of accretion and dispersal of circumstellar material and that
a star's position in the HR diagram {\it is not} primarily age
dependent. (ii) The scatter in the luminosities caused by the nuisance
sources discussed in section~\ref{lumspread} is so large that it erases
the expected age-dependent decrease in the fraction of stars exhibiting
accretion or disk signatures. For the reasons discussed in
section~\ref{lumspread} I regard this latter possibility as
unlikely. In either case (i) or (ii) it would mean that the HR diagram
could not be used to claim a large age spread or to estimate the star
formation history.

\section{Episodic accretion -- a possible explanation}
\label{interpret}

The idea that early accretion could alter a PMS star's
position in the HR diagram and make it appear older have been around
for some time (e.g.  Mercer-Smith, Cameron \& Epstein 1984; Tout et
al. 1999). Recently it has been realised (e.g. by Enoch et al. 2009)
that accretion onto very young stars may be transient or episodic, with
very high accretion rates ($\sim 10^{-4}\,M_{\odot}$\,yr$^{-1}$)
occurring for brief periods of time ($\sim 100$\,yr). ``Episodic
accretion'', which would take place during the early class I T-Tauri
phase, has been modelled by Vorobyov \& Basu (2006) and its
consequences for the PMS HR diagram are explored by Baraffe, Chabrier \&
Gallardo (2009). They find that if the accreted energy is efficiently
radiated away, then a short phase of rapid accretion compresses the PMS
star, leading to a smaller radius and lower luminosity. The star will
not relax back to the configuration predicted by non-accreting models
for a thermal timescale ($\simeq 20$\,Myr for the PMS stars I am
discussing), and hence interpreting the HR diagram using
non-accreting models would lead to erroneously large ages. A
distribution of accretion histories in a coeval SFR could lead to a
luminosity spread and the appearance of an age spread. As there may be
no connection between accretion rates in the class I phase and later
accretion as a class II T-Tauri star this could effectively randomise
the ages determined from the HR diagram for young class II and class III PMS stars.

The model may also account for the apparent spin-down of PMS stars with
age and for the small proportion of stars which appear to have
anomalously high Li depletion.  A PMS star with a true age of say
3\,Myr, that had been subjected to relatively slow accretion rates
during the class I phase would have contracted over 3\,Myr from a
larger radius and spun-up significantly.  A coeval PMS star that had
previously accreted at much high rates would already be smaller, less
luminous and appear older, but would be relaxing back to its
equilibrium configuration on a 20\,Myr timescale and so would have
undergone very limited contraction and spin-up (Littlefair et
al. 2011).  The same stars would have smaller radii and higher central
temperatures than their slow-accreting counterparts and could therefore
burn Li more readily (Baraffe \& Chabrier 2010).

\section{Conclusions}

The evidence to date suggests that the luminosity dispersion seen in
the HR diagrams of young SFRs has a significant component that cannot
be attributed to ``nuisance'' sources such as binarity, variability and
accretion. However, attempts to verify the consequent age spreads
implied by the positions of PMS stars in the HR diagram have
mixed success. In particular, the rotation rates of PMS stars and the
fraction of stars showing active accretion or evidence for
circumstellar material within a single SFR do not show the expected
decrease with age. ``Episodic accretion'' potentially resolves this paradox --
a very high rate of accretion during the class I phase could drive
PMS stars out of equilibrium and towards smaller radii and lower
luminosities. A distribution of early accretion rates would effectively
scramble ages determined from the HR diagram for a population of
class II and class III PMS stars.

If this scenario is borne out by further work, then the traditional HR
diagram is a poor tool for estimating the ages of young ($<20$\,Myr)
PMS stars and also perhaps for estimating age-dependent masses. Large
scale survey work may instead have to rely on less precise but
potentially more accurate clocks such as rotation rates or the presence
of circumstellar material, although of course these may not be
universal and could have significant environmental dependencies.

%
%
%
%

%
%
%

\end{document}